\documentclass[nofootinbib,notitlepage,superscriptaddress,10pt,aps,pra,twocolumn]{revtex4-2}

\usepackage{xcolor}
\usepackage{graphicx}  
\usepackage{subfigure}
\usepackage{multirow}
\usepackage{bm}       
\usepackage{amsfonts}  
\usepackage{mathalfa}
\usepackage{amsmath}
\usepackage{mathrsfs}
\usepackage{euscript}
\usepackage{amssymb} 
\usepackage{cancel}
\usepackage{braket}
\usepackage{float}
\usepackage[font=footnotesize]{caption}

\usepackage{enumitem} 
\usepackage{hyperref}
\makeatletter
\def\l@subsubsection#1#2{} 
\makeatother
\hypersetup{
   pdftitle={QG-defects},
   colorlinks=true,
   linkcolor=black,
   urlcolor=blue
}

\newcommand{\be}{\nopagebreak[3]\begin{equation}}
\newcommand{\ee}{\end{equation}}
\newcommand{\ba}{\nopagebreak[3]\begin{eqnarray}}
\newcommand{\ea}{\end{eqnarray}}

\begin{document}

\title{Dark energy genesis: modeling dissipative effects in primordial cosmology}
\author{Pietro Pellecchia}
\affiliation{Dipartimento di Fisica Ettore Pancini, Università di Napoli “Federico II”,
Complesso Univ.\ Monte S.\ Angelo, I-80126 Napoli, Italy}
\email{pietro.pellecchia2@unina.it}
\affiliation{INFN, Sezione di Napoli, Complesso Univ.\ Monte S.\ Angelo, I-80126 Napoli, Italy}
\author{Alejandro Perez}
\email{perez@cpt.univ-mrs.fr}
\affiliation{Aix Marseille Univ, Université de Toulon, CNRS, CPT, Marseille, France}
\author{Salvatore Ribisi}
\affiliation{School of Physics and Astronomy, Beijing Normal University, Beijing 100875, China.}
\email{salvatore.ribisi@icloud.com}

\begin{abstract}  
In various approaches to quantum gravity, spacetime geometry is understood to emerge from more fundamental discrete structures at the Planck scale. 
As sometimes posited, their presence could lead to dissipative effects in the smooth effective sector. 
In this paper, we develop the idea of non-conservation in gravity, by introducing an effective cosmological model within unimodular gravity, in which a varying cosmological constant arises as a consequence of dissipation.
We show that this requires to incorporate hidden degrees of freedom---termed \textit{quantum gravity defects}---that act as an effective bath for the matter fields.
To illustrate the viability of the framework, we study the case of an Ohmic bath inspired by the Caldeira-Leggett model for Brownian motion, leading to a diffusion equation for the matter energy density. 
The results show that, starting from a primordial universe with no dark energy, dissipation can account for the generation of a small positive cosmological constant.
\end{abstract}

\maketitle

\section{Dissipation in gravity}\label{sec: intro}
The current understanding of gravity, based on Einstein's general theory of relativity (GR), is formulated within the framework of classical physics.
It is then common to consider GR as a smooth-field effective description of space and time which at a fundamental level must present all the characteristics of quantum physics. 
In other words, general relativity should represent the low-energy limit of a more fundamental quantum theory of gravity, with suitable new degrees of freedom, evolving within a full theory of quantum gravity (QG).

However, although QG is currently the research goal of a lively branch of theoretical physics, the emergence of a classical spacetime from any quantum gravity model remains an open issue. 
If spacetime is interpreted as a sort of macroscopic phase resulting from the collective behavior of more elementary constituents which are fundamentally discrete, it is reasonable to expect departures from the standard kinematics of matter at sufficiently high energies. 
It has been proposed that, in such a picture, one particular effect would necessarily arise at the phenomenological level: dissipation~\cite{Liberati:2013usa}.\vspace{2mm}


It is important to note that this hypothesis acquires phenomenological relevance only if the characteristic scale of quantum gravity, say the Planck length $l_p$, acts as a smooth onset scale; namely, if the distinctive effects of quantum gravity begin to manifest themselves already at scales larger, or even much larger, than $l_p$. Such a possibility would not be surprising, as many macroscopic physical systems exhibit precisely this type of smooth onset.

Consider, for instance, Brownian motion in fluids. The chaotic motion of pollen grains suspended in a fluid is caused by collisions with its fundamental microscopic constituents, the molecules, even though the observable effects appear at length scales much larger than the molecular scale. This naturally leads us to speculate that the Planck scale could play an analogous role for spacetime physics, just as the molecular scale does for fluids.

Moreover, the fluctuation-dissipation theorem tells us that Brownian motion is intrinsically related to dissipation~\cite{Kubo1966}. Once again, it is conceivable that an analogous mechanism may be at work in gravity.

Let us synthesize the above considerations in what we might call the \textit{Brownian} hypothesis of quantum gravity: the presence of more fundamental high-energy degrees of freedom, which are not captured by GR's smooth-field approximation of spacetime, should manifest already in some low-energy regime in the form of dissipation and related effects, just as the presence of molecules generates dissipation and related effects in fluids.\vspace{2mm}

We also remark that dissipation is not uncommon in the literature of quantum gravity.
In fact, effects of this nature can arise naturally in different QG scenarios, such as causal set theory~\cite{Dowker2004, Philpott2009} and group field theory~\cite{Jercher:2023kfr, Calcinari:2026hbm}.
These results further motivate the construction of effective descriptions that allow such effects to be characterized in terms of observable quantities and assessed through experimental constraints.

At the classical level, for instance, the gravity theory that emerges as the simplest modification of GR accommodating dissipative phenomena is unimodular gravity (UG)~\cite{Josset:2016vrq}. 
In this framework, dissipation sources a dynamical cosmological constant, yielding phenomenologically rich deviations from the standard $\Lambda$CDM model. 
This mechanism has been extensively explored in recent years~\cite{Perez:2017krv, Perez:2018microscopic, Garcia-Aspeitia:2019yni, Corral:2020lxt, PerezSudarsky2021, Perez:2020cwa, Amadei:2021aqd, Leon:2022kwn, Landau:2022diffusion, Cruz:2024inn, Chakraborty:2024vqa, Bengochea:2025ldo, Kashyap:2026ivg}; notably, under suitable assumptions, unimodular dissipation has been shown to potentially alleviate major cosmological tensions~\cite{Perez:2017krv, Perez:2020cwa}, and account for inflation and the formation of cosmic structures~\cite{Amadei:2021aqd, Leon:2022kwn, Bengochea:2025ldo}.

{ At the quantum level, 
it has been argued \cite{Perez:2014xca, Amadei:2019wjp, Perez:2022jlm, Perez:2023ugg} that the microscopic degrees of freedom responsible for dissipative effects in the low-energy effective description may also provide a natural resolution of Hawking's information puzzle \cite{Hawking:1976ra}. 
This possibility relies on the fact that the separation of scales required for the applicability of the effective field theory paradigm breaks down in the context of black hole formation and evaporation \cite{samiyo}. }

However, existing models of dissipation in UG are largely phenomenological, often relying on \textit{ad hoc} ansätze rather than deriving from fundamental degrees of freedom. 
To bridge this gap, we propose an effective cosmological framework based on the inclusion of hidden degrees of freedom, aiming to provide a more rigorous and microscopic realization of the dissipation paradigm introduced in Ref.~\cite{Josset:2016vrq}.\vspace{2mm}

The paper is organized as follows.
In \autoref{sec: UG}, we introduce the framework of dissipative unimodular gravity and focus on its cosmological setting.
In \autoref{sec: defects}, we show that dissipation requires an interaction of matter with additional degrees of freedom, and we identify the minimal conditions under which such interaction leads to a dissipation equation consistent with UG.
In \autoref{sec: ohmic}, we illustrate the viability of the formalism by studying the particular model of Ohmic dissipation and show that it predicts the generation of a positive cosmological constant.
Finally, \autoref{sec:conclusions} summarizes our conclusions and the appendix contains some further details 
concerning the nature of the noise associated with the Ohmic bath.

\section{Dissipative unimodular gravity}\label{sec: UG}
We start by observing that including dissipation within a metric theory of gravity immediately rules out GR, as Einstein's field equations enforce, by construction, the local conservation of the stress-energy tensor.

The simplest modification of general relativity capable of accommodating dissipative effects is unimodular gravity.
Originally introduced by Einstein himself~\cite{Einstein1919}, UG can be obtained by adding a constraint to the action that fixes the volume density to a given background volume form~\cite{Alvarez:2023utn, Bengochea:2023dep}.\footnote{More precisely, UG has been rediscovered several times via different derivations, e.g., by restricting the variations of the Einstein-Hilbert action to volume-preserving diffeomorphisms~\cite{CarballoRubio2022}, or by gauge-fixing Weyl-transverse gravity~\cite{Alonso-Serrano:2022rzj}.}

The resulting field equations correspond to the trace-free part of the standard Einstein equations:
\begin{equation}
    R_{\mu \nu}-\frac{1}{4} R\, g_{\mu \nu}=8 \pi G\left(T_{\mu \nu}-\frac{1}{4} T\, g_{\mu \nu}\right) \,.
\end{equation}
Defining the diffusion current 
\begin{equation}
    J_{\mu} \equiv8 \pi G \, \nabla^{\nu} T_{\mu \nu} \, ,
\end{equation}
and using the UG integrability condition\footnote{This condition can either be required as an additional assumption, or be derived automatically by requiring that the matter action is invariant under the symmetries of the theory: the volume-preserving diffeomorphisms~\cite{Josset:2016vrq,CarballoRubio2022}.} $dJ=0$ in combination with Bianchi identities, one obtains
\begin{equation}\label{UG equations with lambda}
    R_{\mu \nu}-\frac{1}{2} R\, g_{\mu \nu}+\underbrace{\left[\Lambda_{0}+\int_{\ell} J\right]}_{\Lambda} g_{\mu \nu}=8 \pi G\, T_{\mu \nu}\,,
\end{equation}
where $\Lambda_0$ is a constant of integration and $\ell$ is a one-dimensional path from some reference event.\footnote{The independence of $\Lambda$ on the choice of path $\ell$ is guaranteed by the integrability condition $dJ=0$.}

As Eq.~\eqref{UG equations with lambda} highlights, this generates a term $\Lambda$ which can be interpreted as an effective cosmological constant that is inherently dynamical.
In fact, it depends on the spacetime point via the cumulative effect of the diffusion current $J$~\cite{Josset:2016vrq}.

In the conservative limit $J = 0$, the field equations of standard GR with a cosmological constant $\Lambda_0$ are identically recovered, a property originally noted in Ref.~\cite{Einstein1919}. 
Conversely, when a non-conservation mechanism is active, we enter the regime of dissipative unimodular gravity, which therefore emerges as a mild generalization of GR where dissipative effects uniquely affect the dark energy sector.\vspace{2mm}

Let us now consider the Friedmann-Lemaître-Robertson-Walker (FLRW) metric in the spatially flat case,
\begin{equation}\label{FRW coco}
    d s^2=-d\tau^2+a^2(\tau)\left(d r^2+r^2d\Omega^2\right)\,.
\end{equation}
The cosmological principle restricts the effective $\Lambda$ to be a function of time alone.
Consequently, the Friedmann and continuity equations for UG, derived from Eq.~\eqref{UG equations with lambda}, read
\begin{subequations}\label{eqs9}
\begin{align}
      &3 H^{2}=8 \pi G \rho+\Lambda      \,,\label{bohhhh} \\  
      &\rho'+3 H \left(\rho+p\right) =-\frac{\Lambda'}{8 \pi G} \,,\label{continuity}
\end{align}
\end{subequations}
where $\rho$ and $p$ are the energy density and pressure of the cosmic matter fields, $H\equiv a' / a$ is the expansion rate of the universe, and the prime denotes derivatives with respect to the cosmic time $\tau$. 

Putting together the equations in \eqref{eqs9}, one also finds the Raychaudhuri equation: 
\begin{equation}\label{raycha}
     H'=-4 \pi G \left(\rho+p\right)\,. 
\end{equation} 
Eqs.~\eqref{eqs9}--\eqref{raycha} govern the dynamics of unimodular cosmology. 
When the energy-diffusion current vanishes, $\Lambda$ reduces to a constant $\Lambda_0$, recovering the standard $\Lambda$CDM model. 
Thus, unimodular cosmology acts as a generalization of the standard framework, introducing modifications in the dark energy sector only in the presence of a dissipation process~\cite{Amadei:2021aqd}.\vspace{2mm}

In closing this section, we emphasize that the three cosmological equations \eqref{eqs9}--\eqref{raycha} constitute a system of only two independent differential equations, for three unknown functions of cosmic time: $a(\tau)$, $\rho(\tau)$, and $\Lambda(\tau)$. 
To solve the system, the precise mechanism of energy non-conservation---leading to a time dependence of $\Lambda(\tau)$---must be specified. 
Therefore, we must supplement the dynamics with a third relation for $\Lambda$, the so-called diffusion equation, typically introduced in the literature as a phenomenological ansatz.

\section{The bath of quantum gravity defects}\label{sec: defects}
In this section, we develop the Hamiltonian formalism for unimodular cosmology.
{Let us write the metric~\eqref{FRW coco} as}
\begin{equation}\label{FRW unimodular lapse}
    d s^2=-N^2dt^2+a^2(t)\left(d r^2+r^2d\Omega^2\right)\,,
\end{equation}
where $t$ is a generic time coordinate, and the choice of the lapse function $N$ specifies the time gauging.

As previously mentioned, the action of unimodular gravity can be written as the Einstein-Hilbert action supplemented with the unimodular constraint $\sqrt{|g|}=1$, i.e.
\begin{equation}\label{action generic ug}
    S_G=\kappa \int \left(\sqrt{|g|} R+\lambda(\sqrt{|g|}-1)\right)d^4x\,,
\end{equation}
where $\kappa\equiv 1/16\pi G$.
Specializing to the metric~\eqref{FRW unimodular lapse}, one finds
\begin{equation}
    S_G=\kappa V_0 \int\left(-6 \frac{a \dot{a}^2}{N}+\lambda\left(N a^3-1\right)\right) d t\,,
\end{equation}
where total derivative terms have been eliminated, the 3-volume of a fiducial cell $V_0 \equiv \int d x^3$ has been introduced, and the dot denotes derivatives with respect to time $t$. 

Varying the action with respect to the multiplier $\lambda$ yields the unimodular constraint in the FLRW background, which fixes the lapse function to $N = 1/a^3$. This choice naturally identifies $t$ as the unimodular time variable, related to the cosmic time via $ dt= a^3 d \tau$.
Substituting this constraint back into the action leads to
\begin{equation}\label{azione point particle}
  S_G=-\frac{2}{3} \kappa\int  \dot{v}^2 \,  V_0 \,d t\,,
\end{equation}
where we have introduced the volume variable $v \equiv a^3$.

In this parameterization, the cosmological dynamics mimics that of a free non-relativistic particle. 
The phase-space description is completed by introducing the momentum density conjugate to the volume $v$, which reads
\begin{equation}\label{momentum densities pv}
      p_v=-\frac{4}{3}\kappa \,\dot v \, .
\end{equation}
Since $\dot v=3\,a'/a$, we emphasize that in this parametrization the phase-space variables are essentially given by the spatial volume and the Hubble expansion rate.\vspace{2mm}

To study the full dynamics, let us introduce a scalar field $\phi$ with a generic potential $V(\phi)$ as a matter model.
The corresponding action is given by
\begin{equation}
    S_M =- \int d^4x\sqrt{|g|}\left(\frac{1}{2}\nabla_\alpha \phi \nabla^\alpha \phi+V(\phi)\right) \, .
\end{equation}
Using Eq.~\eqref{azione point particle} for the gravitational sector, the standard action $S\equiv S_G+S_M$ reads
\begin{equation}\label{azione incompleta calcolo}
    \begin{aligned}
     S= \int\left(-\frac{2}{3}\kappa \, \dot{v}^2+\frac{1}{2} v^2 \dot{\phi}^2- V(\phi)\right) V_0 \, d t\,.
    \end{aligned}
\end{equation}
The conjugate momentum density of the scalar field is $p_\phi =  \dot{\phi}\,v^2$. 
Consequently, factoring out the arbitrary fiducial volume, the total Hamiltonian density $\mathcal{H}$ of unimodular cosmology can be cast as
\begin{equation}\label{hamiltonian geo+mat}
    \mathcal{H}=\underbrace{\left(-\frac{3p_v^2}{8\kappa}\right)}_{\mathcal{H}_{\text{G}}}+\underbrace{\left(\frac{p_\phi ^2 }{2  v^2} + V(\phi)\right)}_{\mathcal{H}_{\text{M}}}\, ,
\end{equation}
where the quantities appearing in the matter Hamiltonian $\mathcal{H}_{\text {M}}$ are related to the energy density and pressure of the matter field by 
\begin{equation}\label{Hm e rho}
    \begin{aligned}
       &\rho=\frac{p_\phi ^2 }{2  v^2} +V(\phi)\, , \\
       &p=\frac{p_\phi ^2 }{2  v^2} -V(\phi) \, .
  \end{aligned}   
\end{equation}

At this point, we note that defining the quantity
\begin{equation}\label{lambda def}
    \Lambda_0 \equiv -8 \pi G \, \mathcal{H} \, ,
\end{equation}
and using Eqs.~\eqref{momentum densities pv} and~\eqref{Hm e rho}, one recovers the Friedmann equation~\eqref{bohhhh} written in terms of unimodular time derivatives, that is
\begin{equation}\label{friedman eq from hamilton}
     3 a^4\dot a^2=8 \pi G \, \rho+\Lambda_0 \,,
\end{equation} 
where the quantity $\Lambda_0$ is the cosmological constant.
This provides a clear-cut interpretation for the total energy $\mathcal{H}$ in unimodular cosmology:\footnote{As shown in~\cite{Alonso-Serrano:2022rzj,Odak:2026uac}, this does not rely on any specific cosmological setting, but is a generic feature of theories with a background volume form like unimodular and Weyl-transverse gravity.} 
it is in direct correspondence with the value of the cosmological constant which, in contrast to general relativity, is a dynamical quantity.

Nevertheless, since the Hamiltonian~\eqref{hamiltonian geo+mat} does not explicitly depend on time, $\Lambda_0$ is strictly conserved, precluding the description of dissipative phenomena. 
In other words, the standard $\mathcal{H} = \mathcal{H}_{\text{G}} + \mathcal{H}_{\text{M}}$ alone cannot account for energy non-conservation scenarios. 
We conclude that implementing dissipation in unimodular cosmology requires a suitable extension of the total Hamiltonian.\vspace{2mm}

Building on these considerations, we extend the Hamiltonian in Eq.~\eqref{hamiltonian geo+mat} to a more general total Hamiltonian:
\begin{equation}\label{total hamiltonian mat-geo}
    \mathcal{H}_T = \mathcal{H}_G + \mathcal{H}_M + \mathcal{H}_D \,,
\end{equation}
where we assume that energy can flow between the standard sector (matter and geometry) and a hidden sector described by $\mathcal{H}_D$.
In this framework, diffusion stems from the interaction between the standard degrees of freedom and the hidden ones. 
Consequently, the bare cosmological constant $\Lambda_0\equiv -8\pi G\,\mathcal{H}_T$ can acquire corrections $\delta \Lambda$ from the hidden sector and---depending on the nature of the hidden sector---become dynamical. 
{As we shall show in full detail in \autoref{sec: ohmic}, indeed}, by virtue of total Hamiltonian conservation one obtains an effective dark energy term $\Lambda=\Lambda_0+\delta \Lambda$ that increases when energy flows from $\mathcal{H}_{G/M}$ into $\mathcal{H}_D$, and decreases when the flow is reversed.

This new paradigm of diffusion is no surprise after all.
As discussed previously, effective energy non-conservation traces back to fundamental physics, specifically to the presence of unresolved quantum gravitational degrees of freedom. 
What we understand as ‘lost energy’ is certainly not lost, but rather flows into these fundamental degrees of freedom that are not accessible to the effective description of a smooth manifold: this is the ultimate reason why, in the classic regime, this manifests as effective violations of energy conservation.

It is important to stress that the total system we study is, strictly speaking, a toy model. 
The hidden degrees of freedom we use are not meant to describe the true nature of the Planckian constituents themselves (which remains unknown). 
They merely illustrate in a simple manner how deviations from a perfectly smooth classical geometry could affect the coarse grained dynamics. 
In this spirit, we refer to these  as quantum gravity defects (QG-defects).\vspace{2mm}

In summary, the Hamiltonian framework~\eqref{total hamiltonian mat-geo} admits diffusion in unimodular cosmology via energy exchange between the standard sector and QG-defects.
The natural next step is to determine the conditions under which this provides a diffusion mechanism consistent with unimodular gravity.  
Specifically, we require that $\mathcal{H}_D$ leads to Hamilton's equations reproducing not only a diffusion equation of the form $\Lambda'=f(\Lambda,\rho,a)$, but also the other fundamental equations of unimodular cosmology, namely Eqs.~\eqref{eqs9}--\eqref{raycha}.
The simplest choice compatible with this requirement is to demand that $\frac{\partial \mathcal{H}_D}{\partial v} = \frac{\partial \mathcal{H}_D}{\partial p_v} = 0$, i.e., defects that do not couple to the low-energy geometric degrees of freedom.

{We note that arbitrarily modifying the action once it has been written in terms of symmetry-reduced variables adapted to the FLRW framework is a dangerous procedure. In particular, it can lead to arbitrary violations of general covariance if the resulting action is regarded as being derived from a four-dimensional action principle. The conditions specified in the previous paragraph ensure that the modification breaks general covariance only down to the subgroup of four-dimensional volume-preserving diffeomorphisms, which, as mentioned above, is compatible with the unimodular gravity description advocated here.}

We conclude that the simplest modeling of dissipation in unimodular cosmology requires introducing hidden degrees of freedom, the defects, where the hidden Hamiltonian couples exclusively to the matter fields. In the following section, we explore a specific model for this hidden Hamiltonian and examine its impact on cosmological evolution.

\section{The toy model of Ohmic diffusion}\label{sec: ohmic}
Let us consider the same cosmological setting of \autoref{sec: defects}, where a scalar field $\phi$ with generic potential $V(\phi)$ is minimally coupled to unimodular gravity, corresponding to the action displayed in Eq.~\eqref{azione incompleta calcolo}.
As we have proven, diffusion requires such action to be supplemented with $S_D$, the action of the defects.

To illustrate the framework, we consider the specific case of an `Ohmic' bath. 
This setup directly generalizes the Caldeira-Leggett model for Brownian motion~\cite{CaldeiraLeggett1983}, a standard paradigm for modeling dissipation and diffusive processes in both quantum and complex systems~\cite{CalzettaHu2008, BreuerPetruccione2002, Weiss2021, Fleming:2010bow}.\vspace{2mm}

We model the defects as a bath of harmonic oscillators linearly coupled to the matter field\footnote{We emphasize that these oscillators are harmonic with respect to unimodular time $t$ rather than to cosmic time $\tau$. This ensures the action is compatible with the symmetries of unimodular gravity.
{This is motivated by the idea that spacetime granularity is, in a way, responsible for the emergence at low energies of the preferred unimodular volume structure~\cite{Anderson:1971pn}. Such physical input opens a channel for the diffusive effects that we seek to characterize with our model.}}
\begin{equation}
    S_D=  \int dt\, V_0\left[ \sum_{\alpha} \left(\frac{1}{2}\dot{Q}_\alpha^2-\frac{1}{2} \omega_\alpha^2 Q_\alpha^2\right)-\sum_{\alpha} b_\alpha(t) \phi \,Q_\alpha\right]\,,
\end{equation}
where the $Q_\alpha$ are the hidden degrees of freedom associated to the QG-defects, $b_\alpha(t)$ are the coupling functions, and the sum runs over all defects contained in the fiducial cell.
The conjugate momentum densities are $P_\alpha= \dot{Q}_\alpha$, hence, the total Hamiltonian reads
\begin{equation}\label{hidden hamiltonian mat-hid}
    \begin{aligned}
         \mathcal{H}_T=&-\frac{3p_v^2}{8\kappa}+\frac{p_\phi ^2 }{2 v^2}+\frac{1}{2}M_0^2\phi^2+ \\
         &+\sum_{\alpha} \left(\frac{P_\alpha^2}{2 }+\frac{1}{2} \omega_\alpha^2 Q_\alpha^2\right)+\sum_{\alpha} b_\alpha(t) \,\phi \,Q_\alpha\,,
    \end{aligned}
\end{equation}
where we are taking a potential $V(\phi)=\frac{1}{2}M_0^2\phi^2$, i.e.\ a bare mass $M_0$ for the scalar field.
As we shall show, this ensures a finite effective mass in the Ohmic limit~\cite{CalzettaHu2008}.

We now proceed as follows: we solve the Hamilton's equation for the defects, that is 
\begin{equation}\label{eq for Q}
    \ddot{Q}_\alpha+\omega_\alpha^2 Q_\alpha=-b_\alpha \phi \, ,
\end{equation}
setting initial conditions at a given time $t_d= 0$, and then substitute the solutions in the equation for the field $\phi$.

Following standard procedures for Ohmic baths, we assume that the interaction is switched on adiabatically at $t_d$ and rapidly settles to a constant value, implying $b_\alpha(0)=0$ and $\dot{b}_\alpha(t)=0$ for any subsequent positive time. 
{ As explained in Ref.~\cite{CalzettaHu2008}, further assuming that the field initially satisfies $\phi(0)=0$ allows the solution of Eq.~\eqref{eq for Q} to be written as}
\ba\label{solution defects}
 &&       Q_\alpha(t)= \,\,Q_\alpha(0) \cos \left(\omega_\alpha t\right)+\frac{\dot{Q}_\alpha(0)}{ \omega_\alpha} \sin \left(\omega_\alpha t\right)+\nonumber \\ && -\frac{b_\alpha \phi(t)}{\omega_\alpha^2}
        +\frac{1}{\omega_\alpha^2} \int_0^t \cos \omega_\alpha\left(t-t^{\prime}\right) \dot{\phi}\left(t^{\prime}\right) b_\alpha d t^{\prime} \, .
   \ea
Substituting this into the equation of motion for $\phi$ yields:
\ba \label{EoM for phi}
   &&  v^2 \ddot{\phi}(t)+2v\dot{v} \dot{\phi}(t) +\phi(t) \left(M_0^2-\beta(0) \right)+ \nonumber \\ && \ \ \ \ \ \  +\int_0^t \beta\left(t-t^{\prime}\right) \dot{\phi}\left(t^{\prime}\right) d t^{\prime}=\xi(t)\,,
\ea
where we have defined the \textit{damping function}  
\begin{equation}\label{auxxiliary function}
   \beta\left(t-t^{\prime}\right) \equiv \sum_\alpha  \frac{b_\alpha^2}{ \omega_\alpha^2}\cos \omega_\alpha\left(t-t^{\prime}\right)\, ,
\end{equation}
and the \textit{noise function} 
\begin{equation}\label{noise}
\!\! \!\!     \xi(t) \equiv -\!\! \sum_\alpha b_\alpha \!\!  \left[ Q_\alpha(0) \cos \left(\omega_\alpha t\right)+\frac{\dot{Q}_\alpha(0)}{ \omega_\alpha} \sin \left(\omega_\alpha t\right) \right] .
\end{equation}
With no further assumptions, the matter field is coupled to all harmonic oscillators non-locally via the integral term in Eq.~\eqref{EoM for phi}.
However, the effect of all these degrees of freedom become local  when the bath of harmonic oscillators is of Ohmic type.\vspace{2mm}

More precisely, the Ohmic interaction is defined by requiring that the damping function defined in Eq.~\eqref{auxxiliary function} be such that
\begin{equation}\label{ohmic}
    \beta\left(t-t'\right)=\bar\beta \, m_p\hspace{0.5mm}\delta(t-t')\,,
\end{equation}
where $\delta(t-t')$ is the Dirac delta, $m_p$ is the Planck mass, and $\bar\beta$ is a dimensionless parameter scaling the characteristic energy scale of the interaction.\footnote{Throughout this work, we adopt natural units $c=\hbar=1$, for which the Planck mass is related to Newton's constant as $m_p = G^{-1/2}$.}
The crucial point of the Ohmic condition~\eqref{ohmic} is that it turns the integral term in Eq.~\eqref{EoM for phi} into a damping term, as is typical of dissipative processes~\cite{CalzettaHu2008}.
In addition (and most importantly), the Ohmic condition renders the dissipative effect of the bath local.

However, this choice manifestly leads to a divergence in $\beta(0)$ due to the Dirac delta, which must be absorbed by renormalizing the bare mass parameter $M_0$.
Namely, one can formally regularize the diverging term by imposing $M_0^2-\beta(0) = M^2$ for some non-negative finite mass $M$, so that Eq.~\eqref{EoM for phi} reduces to
\begin{equation}\label{eom34}
    v^2 \ddot{\phi}(t)+2\, v\, \dot{v}\, \dot{\phi}(t) + M^2\phi(t)+ \bar\beta \,m_p \dot{\phi}(t)=\xi(t)\,.
\end{equation}
In other words, if the interaction with the defects is Ohmic, the matter Hamilton's equation becomes a Langevin-like equation, and the matter field acquires a renormalized finite mass $M$.

This can also be seen at the Hamiltonian level by evaluating $\mathcal{H}_T$ on-shell for $Q_\alpha$.
{Substituting Eq.~\eqref{solution defects} into Eq.~\eqref{hidden hamiltonian mat-hid} yields
\begin{equation}\label{H on shell}
     \mathcal{H}_T =
     -\frac{3p_v^2}{8\kappa} + \frac{p_\phi^2}{2 v^2} + \frac{1}{2} M_0^2 \phi^2-\delta V + \frac{\delta\Lambda}{8\pi G} \, ,
\end{equation}
where we introduced the counterterms $\delta V\equiv\frac{1}{2}\beta(0)\phi^2$ and 
\begin{equation}\label{28}
    \delta \Lambda\equiv {\cal E}_0-\int_0^t \xi(t') \dot{\phi}(t') dt' + \frac{1}{2} \bar{\beta} \, m_p \int_0^t \dot{\phi}^2\!(t')\,dt'  \, ,
\end{equation}
with ${\cal E}_0$ being the initial energy density of the defects (see appendix).

This result clarifies how the hidden sector backreacts on both the matter and dark energy dynamics. 
On the one hand, the bare matter potential is corrected by the counterterm $\delta V$, bringing the matter Hamiltonian into the form of a scalar field with mass $M$, and
 matter density and pressure given by
\begin{equation}\label{effective matter density and pressure after param}
  \begin{aligned}
       &\rho=\frac{1}{2} v^2 \dot{\phi}^2+\frac{1}{2}M^2 \phi^2\, , \\
       &p=\frac{1}{2} v^2 \dot{\phi}^2-\frac{1}{2}M^2\phi^2 \, .
  \end{aligned}   
\end{equation}
On the other hand, the strictly conserved bare cosmological term $\Lambda_{0}\equiv -8\pi G\,\mathcal{H}_T$ is now corrected by the counterterm $\delta \Lambda$ as
\begin{equation}\label{final effective lambda}
     \Lambda \equiv \Lambda_{0} + \delta \Lambda \, .
\end{equation}
This plays the role of an effective dynamical dark energy term. 
Indeed, multiplying the on-shell Hamiltonian~\eqref{H on shell} by $-8\pi G$ and substituting the renormalized matter density~\eqref{effective matter density and pressure after param}, one finds that the Friedmann equation \eqref{friedman eq from hamilton} gets modified into 
\begin{equation}\label{effective friedman unimodular}
     3 a^4\dot a^2 = 8 \pi G \, \rho + \Lambda \, .
\end{equation}

As discussed in \autoref{sec: UG}, the unimodular cosmological equations~\eqref{eqs9}--\eqref{raycha} are underdetermined; the dynamics cannot be solved without specifying a diffusion mechanism.
Using our QG-defect formalism, we now demonstrate that Ohmic interaction naturally yields a well-defined diffusion equation that fully determines the dynamics.
The unimodular time derivative of $\Lambda$ in Eq.~\eqref{final effective lambda} straightforwardly gives
\begin{equation}\label{diffusion equation mat noise}
   \frac{\dot{\Lambda}}{8 \pi G}=\bar\beta\, m_p\dot{\phi}^2-\dot{\phi}\, \xi(t)\,.
\end{equation}

From now on, we assume that the noise $\xi(t)$ can be neglected when solving for the dynamics. 
This hypothesis ultimately amounts to an assumption about the initial conditions: namely, that the bath is sufficiently unexcited at the big bang. 
This is consistent with the requirement that the hidden sector produces only small deviations from standard cosmological phenomenology in the long-time asymptotic (we prove and comment on this in the appendix).  
The special nature of such initial conditions is another instance of the need of a low entropy big bang in cosmology~\cite{Penrose1979}.

At this point, we switch to cosmic time derivatives via $dt=a^3d\tau$. 
Recalling the definition $v = a^3$ and using Eq.~\eqref{effective matter density and pressure after param}, the previous relation yield a diffusion equation:
\begin{equation}\label{diffusion eq}
    \frac{\Lambda'}{8 \pi G}= \bar\beta \, m_p (\rho+p) \, a^{-3},
\end{equation}
alongside the expected unimodular continuity equation:
\begin{equation}\label{emergent conitnuity}
    \rho^{\prime}+3 \frac{a'}{a} (\rho+p)+\frac{\Lambda'}{8 \pi G}=0 \, .
\end{equation}
This confirms that the standard unimodular cosmology is completely recovered, but with a crucial upgrade: an emergent diffusion equation explicitly driven by the interaction with the underlying QG-defects.

Together with the Friedmann equation~\eqref{effective friedman unimodular} and continuity equation~\eqref{emergent conitnuity}, this relation forms a closed set of dynamical equations, setting the stage for exploring the cosmological phenomenology implied by this dissipative model.\vspace{2mm}

First, it is convenient to introduce the dimensionless diffusion parameter 
\begin{equation}
    \gamma\equiv \frac{2\bar\beta}{a_d^3} \, ,
\end{equation}
where $a_d \equiv a(\tau_d)$ is the scale factor at the onset of dissipation ($t = t_d$).
The diffusion equation~\eqref{diffusion eq} can then be written as:\footnote{Notice that this formulation ensures the diffusion equation depends exclusively on the dimensionless ratio $a/a_d$ rather than the bare scale factor $a(\tau)$, which lacks independent physical meaning. This is consistent with the standard cosmological equations and naturally sets $a(\tau_d) = a_d$ as the initial condition for the subsequent evolution.}
\begin{equation}\label{diffusion equation matter nonoise}
     \frac{\Lambda'}{8 \pi G}=\frac{\gamma}{2} \, m_p \,\left(\frac{a_d}{a}\right)^3 \, (\rho+p)  \, .
\end{equation}
This relation shows that a positive diffusion parameter $\gamma$ induces an energy transfer from the matter sector to the hidden defects, which manifests as a monotonic increase in the effective cosmological constant.

This can be quantified by solving the dynamics perturbatively in $\gamma$.
For illustrative purposes, we restrict our analysis to the massless case $M=0$.\footnote{This amounts to consider a universe filled with stiff matter, i.e.\ with equation of state $p=\rho$. Using $M=0$ instead of $M>0$ simplifies the perturbative resolution, but essentially gives the same overall behavior of the cosmological constant, as can be directly shown numerically or analytically.} 
In this scenario, the complete set of equations reads
\ba\label{set final}
                    &&\,\,\,3\left(\frac{a^{\prime}}{a}\right)^2 = 8\pi G\,\rho+\Lambda\,, \nonumber \\
               &&\,\,\,\rho' + 6\frac{a^{\prime}}{a}\rho + \frac{\Lambda'}{8\pi G} = 0\,, \\
               &&\,\,\,\frac{\Lambda'}{8\pi G}= \gamma\,m_p\,\left(\frac{a_d}{a}\right)^3 \nonumber \rho\,.
     \ea
Imposing the initial conditions $\Lambda(\tau_d)=0$ and $\rho(\tau_d)=\rho_d$, the first-order solution for the effective cosmological constant is readily found to be:
\begin{equation} \label{lambda pertu}
    \Lambda(\tau) \simeq \frac{\gamma\, m_p H_d}{2} \left[ 1 - \frac{1}{\left( 3 H_d \left( \tau - \tau_d \right) + 1 \right)^2} \right]+\mathcal{O}(\gamma^2) \, ,
\end{equation}
where $H_d \equiv \sqrt{\frac{8\pi G}{3}\rho_d}$ is the initial Hubble parameter.
{One can solve the rest of the equations easily using perturbation theory in order to describe the dynamics while the cosmological constant is subdominant. The zeroth order solutions are
$\rho=\rho_d (a^6_d/a^6)+\mathcal{O}(\gamma)$, and the corrections can be integrated from the second line in \eqref{set final} using 
\eqref{lambda pertu}. One can treat the Friedmann equation similarly and obtain that $a(\tau)=a_d \left( 3 H_d \left( \tau - \tau_d \right) + 1 \right)^{1/3}+\mathcal{O}(\gamma)$. There is an intermediate regime where perturbation theory fails and numerical integration is needed. In the long time regime, however, matter dilutes, the asymptotic value of the cosmological constant dominates and the spacetime geometry becomes De Sitter.}

\begin{figure}
    \centering
    \includegraphics[width=0.95\linewidth]{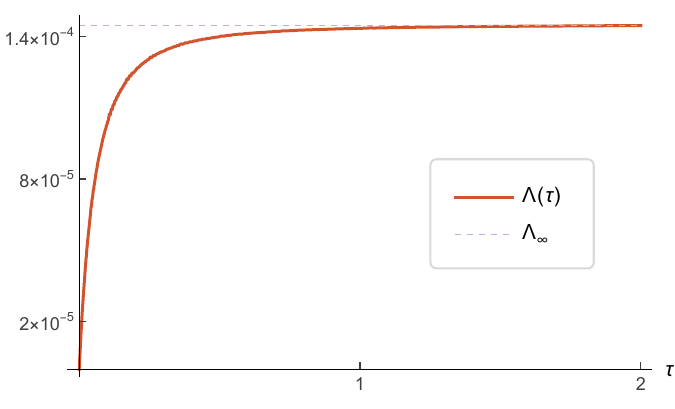}
    \captionsetup{font={footnotesize}}
    \caption{Numerical evolution of the effective cosmological constant $\Lambda(\tau)$ for $\gamma=10^{-4}$, obtained from the set in Eq.~\eqref{set final} with initial conditions $\Lambda_d=0$ and $\rho_d=1$ in Planck units. Dissipative effects generate a cosmological constant which asymptotically converges to the value $\Lambda_\infty$ predicted analytically in Eq.~\eqref{lambda toto} (dashed line).}
    \label{numerics}
\end{figure}

In the late-time limit, $\Lambda(\tau)$ asymptotically approaches the constant value
\begin{equation}\label{lambda toto}
     \Lambda_\infty = \frac{\gamma\, m_p H_d}{2} \, ,
\end{equation}
over a characteristic timescale set by $H_d^{-1}$, as evident from Eq.~\eqref{lambda pertu}.
Consequently, an initially vanishing cosmological constant is dynamically driven to a small, positive asymptotic value by the Ohmic dissipation of matter. 
This analytical prediction is fully corroborated by numerical integration, as shown in \autoref{numerics}.

\section{Conclusions}\label{sec:conclusions}
In this work, we have discussed the emergence of dissipation in gravity as a macroscopic manifestation of the underlying quantum gravitational degrees of freedom. By developing the dissipation paradigm within unimodular gravity, we showed that the effective cosmological constant is sourced by diffusion of the classical matter-geometry system. 

Our canonical analysis reveals that this effectively open dynamics necessitates the existence of a hidden sector.
To formalize this, we introduced a toy model of  quantum gravity defects offering a channel to  the energy exchange between classical matter fields and the granular structure of spacetime.\vspace{2mm}


To illustrate the concrete viability of this framework, we specialized the interaction to an Ohmic bath, inspired by the Caldeira-Leggett paradigm of Brownian motion. 
In a simple toy model with a single scalar field as matter content, we showed that the ultraviolet divergences inherent to this setup can be absorbed via standard mass renormalization. 

This procedure yields a Langevin-like equation of motion for the scalar field, featuring both damping and noise terms, while the field acquires a finite effective mass $M$. 
Consequently, the diffusion equation is analytically derived from an underlying microscopic model rather than being postulated as a phenomenological ansatz.

The framework thus provides a complete set of cosmological equations where deviations from standard dynamics are governed by diffusion. For a small diffusion parameter $\gamma$ {and neglecting the noise $\xi(t)$ (in consistency with the assumption of low initial big bang entropy)}, this system can be solved analytically at the perturbative level. 
Tracking the cosmological evolution reveals that this Brownian-like dissipation dynamically generates a small, positive asymptotic cosmological constant $\Lambda_\infty$ at late times, starting from a universe with initially vanishing dark energy.\vspace{2mm}

This analytical result, fully corroborated by numerical integration, provides a { natural}, microscopically motivated mechanism for the emergence of dark energy as a direct consequence of energy diffusion. 
Ultimately, the bath of quantum gravity defects successfully grounds the phenomenological dissipation paradigm originally proposed in Refs.~\cite{Josset:2016vrq, Perez:2017krv, Perez:2018microscopic}.

%

Note that Eq.~\eqref{lambda toto} can be recast as
$\Lambda_\infty = \gamma \sqrt{{2\pi\rho_d}/{3}}$,
showing that the asymptotic cosmological constant is parametrically tied to the key physical quantities governing the dissipative process: the dimensionless coupling $\gamma$ and the initial energy density $\rho_d$ at the onset of diffusion.

Current observations constrain dark energy to $\Lambda_{\text{obs}} \sim 10^{-120}$ in Planck units. 
In standard general relativity, this tiny value clashes with the Planckian contributions expected from vacuum fluctuations, constituting the core of the notorious cosmological constant problem~\cite{weinberg}. 

Crucially, within the general framework of unimodular gravity---regardless of dissipative effects---vacuum fluctuations do not gravitate, effectively decoupling from the spacetime dynamics~\cite{Alvarez:2023utn,CarballoRubio2022}. 
With the vacuum energy eliminated as a gravitational source, the problem naturally shifts from protecting $\Lambda$ from fine-tuning to identifying the actual physical mechanism responsible for the observed accelerated expansion. 

Dissipative unimodular gravity offers a compelling answer: $\Lambda_\infty$ is dynamically generated by the interaction of matter fields with the  bath of QG-defects and remains closely tied to the fundamental parameters of the dissipation process, as is evident from Eq.~\eqref{lambda toto}.

{ As is the case for many symmetry-reduced models in quantum cosmology, the toy model studied in this work is too simple to provide quantitative information regarding the previous question. However, it offers a novel conceptual perspective that we hope will prove insightful for future investigations, which will require a more detailed understanding of the microscopic theory (quantum gravity) as well as a more realistic modeling of the matter content.  

Finally, and connecting with more fundamental aspects,  diffusion and noise have been treated here in a completely classical way. 
Quantum effects~\cite{Weiss2021, Fleming:2010bow, Fahn:2026laq} like decoherence and localization induced by the microscopic structure are potentially relevant in view of connecting the paradigm proposed in Refs.~\cite{Amadei:2019wjp, Bengochea:2025ldo} for understanding the origin of structure in primordial cosmology. 
We plan to investigate this perspective in the future. }

%

\begin{acknowledgments}
S.R. and P.P.\ thank Federico Greco for helpful discussions.
S.R. is grateful for support from the National Natural Science Foundation of China (Grant No.12275022).
This work falls within the scope of the COST Action CA23130 “Bridging high and low energies in search of quantum gravity”, and WOST (https://withoutspacetime.org), supported
by Grant ID63683 from the John Templeton Foundation
(JTF). The opinions expressed in this work are those of
the author(s) and do not necessarily reflect the views of
the John Templeton Foundation.
\end{acknowledgments}

\appendix
\section{Fluctuation dissipation and the initial state of the bath}\label{appendix}
As in the standard formulation of Ohmic interactions~\cite{CalzettaHu2008}, the bath of harmonic oscillators is initially in an equipartition of probability equilibrium state. In usual cases the state is parametrized by a bath temperature. 
The initial conditions $Q_\alpha(0)$ and $\dot{Q}_\alpha(0)$ are therefore random variables distributed according to    
\begin{equation}\label{app1}
   \begin{aligned}
        &\langle Q_\alpha(0) \rangle = \langle \dot{Q}_\alpha(0) \rangle =\langle Q_\alpha(0) \dot{Q}_{\alpha'}(0) \rangle = 0 \,  , \\
       &\langle \dot{Q}_\alpha(0) \dot{Q}_{\alpha'}(0) \rangle = \omega_\alpha^2\langle Q_\alpha(0) Q_{\alpha'}(0) \rangle =\delta_{\alpha\alpha'} {T_d^4} \, ,
   \end{aligned}
\end{equation}
where ensemble averages are denoted by $\langle \dots \rangle$, 
we are taking the Boltzmann constant $k_B=1$, and we are introducing a primordial temperature $T_d$ of the defects to characterize the stochastic properties of the state (the fourth power coming from dimensional considerations). 

Using these conditions, one finds that the initial energy density of the defects introduced in Eq.~\eqref{28}, that is
\begin{equation}
    \mathcal{E}_D^0\equiv\frac{1}{2}\sum_{\alpha} \left(\dot Q_\alpha(0)^2 + \omega_\alpha^2 Q_\alpha(0)^2\right) \, ,
\end{equation}
is related to the defects temperature by $\langle \mathcal{E}_D^0\rangle=T_d^4$.

Moreover, from Eq.~\eqref{app1} it immediately follows that the expected value of the noise is identically zero, $\langle \xi(t) \rangle = 0$, while using Eq.~\eqref{noise} the noise autocorrelation function is
\begin{equation}
    \langle \xi(t)\xi(t') \rangle = T_d^4\, \beta(t-t') \, ,
\end{equation}
where we used the definition of damping function~\eqref{auxxiliary function}.
This is the formal expression of the classical Fluctuation-Dissipation theorem.
Specializing the result to the Ohmic case, condition~\eqref{ohmic} imposes the noise autocorrelation function to be:
\begin{equation}\label{pietro}
    \langle \xi(t)\xi(t') \rangle = T_d^4 \,\bar{\beta} m_p \delta(t-t') \, .
\end{equation}
Now, equation of motion~\eqref{eom34} in the massless case reads 
\begin{equation}
    v^2 \ddot{\phi} + 2 v \dot{v} \dot{\phi} + \bar{\beta} m_p \dot{\phi} = \xi (t) \, .
\end{equation}
In the late time asymptotic  $\Lambda \sim$ constant, the Friedmann equation in terms of unimodular time gives
\begin{equation}
     \frac{\dot{v}^2}{3} = \Lambda \, ,
\end{equation}
implying $v =\sqrt{3 \Lambda} t$. This allows us to rewrite the diffusion equation as
\begin{equation}
    3 \Lambda t^2 \ddot{\phi} + 6 \Lambda t \dot{\phi} + \bar{\beta} m_p \dot{\phi} = \xi (t) \, ,
\end{equation}
%
%
%
%
%
from which it follows 
\begin{equation}
     \dot{\phi} (t) =  \frac{e^{\frac{\bar{\beta} m_p}{3 \Lambda t}}}{t^2} \left( \int_{0}^t dt'  \frac{e^{- \frac{\bar{\beta} m_p}{3 \Lambda t'}} \xi (t')}{3\Lambda} + K \right) \, .
\end{equation}
with $K$ an integration constant.

At this point, computing the ensemble average
\begin{align} \langle \dot{\phi}^2 \rangle = \frac{2 \rho}{v^2} = \frac{2 \rho}{3 \Lambda t^2}\nonumber \end{align} and 
\begin{equation}
    \langle \xi(t) \dot{\phi} (t) \rangle = \frac{e^{\frac{\bar{\beta} m_p}{3 \Lambda t}}}{3 \Lambda t^2}\int_{0}^t dt' e^{- \frac{\bar{\beta} m_p}{3 \Lambda t'}} \left\langle  \xi(t') \xi (t) \right\rangle 
   = \frac{\bar \beta m_p T_d^4}{3 \Lambda t^2} \, \nonumber ,
\end{equation}
where we used Eq.~\eqref{pietro}.
Replacing the previous in Eq.~\eqref{diffusion equation mat noise}, and demanding the equilibrium condition 
$\langle\dot \Lambda \rangle=0$, we get
\begin{equation}
    \rho  = T_d^4  \, ,
\end{equation}
which implies that radiation will equilibrate with the primordial defect bath at the primordial temperature $T_d$.

{ Replacing this back into the Friedmann equation we find that
\begin{equation}
    \Lambda=\Lambda(\infty)+8\pi \frac{T_d^4}{m_p^2}\,,
\end{equation}
where $\Lambda (\infty)$ is the value of the cosmological constant obtained neglecting the noise.

The calculations of this appendix are expected to reflect the leading effects of the noise in the model when treated as a perturbation of the noiseless equations. On the opposite regime, if $T_d$ is order Planck, then we expect the noise to dominate over the initial radiation distribution and the dynamics to yield a De Sitter expansion with a Planck-scale cosmological constant and radiation in thermal equilibrium with the defects.}

%


\end{document}